\documentclass[12pt,A4paper]{article}
\usepackage{theorem}
\usepackage{makeidx}
\usepackage{amsmath}
\usepackage{amssymb}
\usepackage{epsfig}
\usepackage[sf,small]{titlesec}
\newtheorem{Theorem*}{Theorem*}
\newtheorem{Theorem}{Theorem}

\newtheorem{Definition}{Definition}
\newtheorem{Example}{Example}

\theoremstyle{Remark}

\numberwithin{Proposition}{section}
\numberwithin{Theorem}{section}
\numberwithin{Corollary}{section}
\numberwithin{Definition}{section}
\numberwithin{Example}{section}
\numberwithin{Remark}{section}
\numberwithin{Condition}{section}
\numberwithin{Assumption}{section}
\usepackage{graphics}
\usepackage[latin1]{inputenc}
\begin{document}
\noindent
\renewcommand{\thefootnote}{\fnsymbol{footnote}}
\thispagestyle{empty}

\begin{center} 
{\bf\Large{A note on projections of  Gibbs measures  from a class arising in economic modeling}}\\
\vspace{0.6cm}
M. Hohnisch and O. Kutoviy{\footnote{
Address: M.H.: Department of Economics, University of Bonn, Adenauerallee 24-42, 53113 Bonn, Germany, email:
Martin.Hohnisch@uni-bonn.de;
O.K.: Department of Mathematics, University of Bielefeld, D-33615 Bielefeld, 
email:kutoviy@mathematik.uni-bielefeld.de}}

\vspace{0.3cm}
October 24, 2006 
\end{center}
\vspace{0.2cm}
{\bf{Abstract}}:
A result about projections of  Gibbs measures from a particular class arising in economic modeling is proved.
\vspace{0.5cm}
\section{Introduction}
\renewcommand{\thefootnote}{\arabic{footnote}}
\setcounter{footnote}{0}
Though individual human behavior  undoubtedly is far more complex than the behavior of individual objects
in Physics, in the study of  coordination phenomena in large economic systems a reasonable approach is to restrict attention 
to certain known individual behavioral  regularities of consumers, traders etc which are simple and stable enough to allow 
Statistical-Mechanics-based modeling. Such an approach leaves the state of any single individual random,
 and makes instead the  {\it{collective}} behavior of agents endogenous in a model, given the a-priori known individual behavioral regularities 
which are formalized as conditional probability distributions of individual variables \cite{F}. 
Thus it corresponds to the Dobrushin-Lanford-Ruelle approach to defining Gibbs measures 
on countably-infinite  structures \cite{G}. 

A natural property of  models in  Economics is  that  typically
multiple individual variables of distinct types must be introduced to characterize a {\it{single}} individual economic entity (``agent''), 
and the macroscopic variables of the system are determined by the interaction of all  variables of all agents. 
Taking the statistical approach, one is thus  led to Gibbsian fields with multiple types of variables, with interactions both between
variables of the same type associated to different agents and variables of different types associated to the same agent.
Since not all types of variables are of immediate economic relevance, one  is typically interested in  
projected infinite-volume Gibbs measures corresponding to a given subset of variable types.
In that context, the present note provides a result which for a specific structure of interactions 
simplifies the computation of the projected measure.

\section{The result}
Let $\mathbb{Z}^{d}$ denote the set of agents in a large economy. 
To each agent $i \in \mathbb{Z}^{d}$ there are associated two
variables  $x_i$ and $y_i$ with values in $X$ and $Y,$ respectively. 
For concreteness, we consider the case $X=Y=\mathbb{R}^{n}.$ 
We set $\Omega_{X}=\Omega_{Y}:=(\mathbb{R}^{n})^{\mathbb{Z}^{d}}$
and $\Omega:=\Omega_{X}\times\Omega_{Y}=(\mathbb{R}^{n}\times\mathbb{R}^{n})^{\mathbb{Z}^{d}}.$
 By
$\mathcal{B}(\Omega_{X})=\mathcal{B}(\Omega_{Y})$ and $\mathcal{B}(\Omega)$
we denote the corresponding Borel $\sigma$-algebras of these spaces.
For any $\Lambda\subset \mathbb{Z}^{d}$ we define
$\Omega_{X}^{\Lambda}=\Omega_{Y}^{\Lambda}:=(\mathbb{R}^{n})^{\Lambda}$ and
$\Omega_{\Lambda}:=\Omega_{X}^{\Lambda}\times\Omega_{Y}^{\Lambda}.$
An element of $\Omega$  will be denoted by $x\times y$.

A Gibbs measure $\mu$ on $\Omega$ appropriately represents an equilibrium state  of a large multi-component system with some given
structure of local interactions between them. 
For any finite $\Lambda\subset\mathbb{Z}^{d}$ and  any $\bar{x}\times
\bar{y}\in\Omega$   the corresponding conditional Gibbs measures in finite volumes are of the form
$$\mu_{\Lambda}\left(dx_{\Lambda}\times
dy_{\Lambda}\,|\,\bar{x}\times\bar{y}\right)=
\frac{1}{Z_{\Lambda}(\bar{x}\times\bar{y})}p_{\Lambda}\left(x_{\Lambda}\times
y_{\Lambda}\,|\,\bar{x}\times\bar{y}\right)dx_{\Lambda}dy_{\Lambda},
$$
where
$$
Z_{\Lambda}(\bar{x}\times\bar{y}):=\int_{\Omega_{\Lambda}}p_{\Lambda}\left(x_{\Lambda}\times
y_{\Lambda}\,|\,\bar{x}\times\bar{y}\right)dx_{\Lambda}dy_{\Lambda}
$$
is the so-called partition function, and
$p_{\Lambda}\left(x_{\Lambda}\times y_{\Lambda}\,|\,\bar{x}\times\bar{y}\right)$
the conditional density for variables in $\Lambda$ derived from the $\Lambda$-Hamiltonian given 
the configuration $(\bar{x}\times\bar{y})_{\Lambda^c}.$

Motivated by certain economic models, we consider Gibbs measures from the following class: 
\begin{Definition}
Let $\mathcal{G}_0$ denote the class of Gibbs measures on $\Omega$, whose corresponding conditional
distributions do not depend on condition from $\Omega_{X}$, i.e.
$$p_{\Lambda}\left(x_{\Lambda}\times
y_{\Lambda}\,|\,\bar{x}\times\bar{y}\right)=p_{\Lambda}\left(x_{\Lambda}\times
y_{\Lambda}\,|\,\bar{y}\right)$$
\end{Definition}

\begin{Example}
The following  conditional densities fulfill the condition in the above definition 
(see \cite{H} for the economic motivation behind this particular interaction structure).
$$p_{\Lambda}\left(x_{\Lambda}\times y_{\Lambda}\,|\,\bar{x}\times
\bar{y}\right)=$$
$$=\exp{\left\{-\sum_{i\in\Lambda}x_{i}^{2}-J_{c}\sum_{i\in\Lambda}(x_{i}-y_{i})^{2}-
J_{s}\sum_{\langle
i,\,j\rangle\in\Lambda}(y_{i}-y_{j})^{2}-J_{s}\sum_{\langle
i,\,j\rangle\in\mathbb{Z}^{d}, \ i \in \Lambda, \ j \in \Lambda^c}(y_{i}-\bar{y}_{j})^{2}\right\}},
$$
with a finite $\Lambda\subset\mathbb{Z}^{d}$ and $\langle i,\,j\rangle$ denoting all 
$i,\,j$ such that $|i-j|=1$.
\end{Example}

Since only a subset of  variable-types is of direct
economic relevance, one typically is led to the problem of computing certain projected infinite-volume Gibbs measures.
In the specific context specified above, we obtain the following result about the projected measure.
\begin{Theorem}
Suppose that $\mu\in\mathcal{G}_0$. Then the measure
$\mu_{eff}^{Y}$ which is defined on $\Omega_{Y}$ by
$$
\mu_{eff}^{Y}(A)=\mu(\Omega_{X}\times A),\;\;\;
A\in\mathcal{B}(\Omega_{Y})
$$
will be a Gibbs measure whose corresponding conditional measures in
finite volume $\Lambda\subset\mathbb{Z}^{d}$ are given by
\begin{equation}
\mu_{\Lambda,\,eff}(dy_{\Lambda}\,|\,\bar{y})=\int_{\Omega_{X}^{\Lambda}}\mu_{\Lambda}
(dx_{\Lambda}\times dy_{\Lambda}\,|\,\bar{y})=\label{1}
\end{equation}
$$
=\int_{\Omega_{X}^{\Lambda}}\mu_{\Lambda} (dx_{\Lambda}\times
dy_{\Lambda}\,|\,\bar{x}\times\bar{y}), \ \ \ \bar{x}\times\bar{y}\in\Omega.
$$
\end{Theorem}
%\begin{proof} 
PROOF: Let us consider Gibbs specifications which corresponds
to the measures $\mu_{eff}^{Y}$, for any $\Lambda\subset\mathbb{Z}^{d}$-finite
and $\bar{y}\in\Omega$ given by the probability kernel
\begin{equation}
\pi_{\Lambda}^{Y}(A\,|\,\bar{y})=\int_{A'}\mu_{\Lambda,\,eff}(dy_{\Lambda}\,|\,\bar{y}),\label{2}
\end{equation}
where
$
A':=\left\{y\in\Omega_{Y}^{\Lambda}\,|\,y\times\bar{y}_{\Lambda^{c}}\in
A\right\},$ and  $A\in\mathcal{B}(\Omega_{Y}).$
Equation 1 implies now the following
$$
\pi_{\Lambda}^{Y}(A\,|\,\bar{y})=\int_{A'}\int_{\Omega_{X}^{\Lambda}}\mu_{\Lambda}(dx_{\Lambda}\times
dy_{\Lambda}\,|\,\bar{y})=
$$
$$
=\pi_{\Lambda}(\Omega_{X}\times A\,|\,\bar{y}),
$$
where $\pi_{\Lambda}(\cdot\,|\,\bar{y})$ are the Gibbs specifications 
corresponding to the measure $\mu$. We show that the DLR-equations for the measure $\mu_{eff}^{Y}$ hold.
$$(\mu_{eff}^{Y}\pi_{\Lambda}^{Y})(A)=\int_{\Omega_{Y}}\pi_{\Lambda}^{Y}(A| \ \bar{y})\mu_{eff}^{Y}(d\bar{y})
=\int_{\Omega_{Y}}\pi_{\Lambda}(\Omega_{X}\times
A| \ \bar{y})\mu(\Omega_{X}\times d\bar{y})=
$$
$$
=\int_{\Omega_{X}}\int_{\Omega_{Y}}\pi_{\Lambda}(\Omega_{X}\times
A|\ \bar{y})\mu(d\bar{x}\times d\bar{y}).
$$
But by the DLR-equations for the measure $\mu$  the latter expression is indeed equal to
$$
\mu(\Omega_{X}\times A)=\mu_{eff}^{Y}(A).
$$
%\end{proof} 

\end{document}